\documentclass[apjl]{emulateapj}
\usepackage{apjfonts}

\def\alwaysmath#1{\ifmmode{#1}\else{$#1$}\fi}

\shorttitle{Measuring Sizes of Ultra-Faint Dwarfs}
\shortauthors{Mu\~noz et al.}

\begin{document}

\title{Measuring Sizes of Ultra-Faint Dwarf Galaxies}

\author{R. R. Mu\~noz\altaffilmark{1,2}, 
N. Padmanabhan\altaffilmark{3,2}, \&
M. Geha\altaffilmark{2,3}} 

\altaffiltext{1}{Departamento de Astronom\'ia, Universidad de Chile, Casilla 36-D, Santiago, 
Chile (rmunoz@das.uchile.cl)}
\altaffiltext{2}{Dept. of Astronomy, Yale University, New Haven, CT 06520}
\altaffiltext{3}{Dept. of Physics, Yale University, New Haven, CT 06520}

\begin{abstract}

 The discovery of Ultra-Faint Dwarf (UFD) galaxies in the halo of the
 Milky Way extends the faint end of the galaxy luminosity function to
 a few hundred solar luminosities.  This extremely low luminosity
 regime poses a significant challenge for the photometric
 characterization of these systems.  We present a suite of
 simulations aimed at understanding how different observational
 choices related to the properties of a low luminosity system impact
 our ability to determine its true structural parameters such as
 half-light radius and central surface brightness.  We focus on
 estimating half-light radii (on which mass estimates depend
 linearly) and find that these numbers can have up to 100\%
 uncertainties when relatively shallow photometric surveys, such as
 SDSS, are used.  Our simulations suggest that to recover structural
 parameters within 10\% or better of their true values: (a) the ratio
of the field-of-view to the half-light radius of the satellite must
 be greater than three, (b) the total number of stars, including
 background objects should be larger than $1000$, and (c) the central
 to background stellar density ratio must be higher than $20$.  If
 one or more of these criteria are not met, the accuracy of the
 resulting structural parameters can be significantly compromised.
 In the context of future surveys such as LSST, the latter condition
 will be closely tied to our ability to remove unresolved background
 galaxies. 
 Assessing the reliability of measured structural parameters will 
 become increasingly critical as the next generation of deep wide-field 
 surveys detects UFDs beyond the reach of current spectroscopic limits.

\end{abstract}


\section{Introduction}
\label{sec:intro}

The structural parameters of galaxies are of fundamental importance.
Uncertainties in estimating a galaxy's physical size, central surface
brightness, or total luminosity can quickly propagate into other key galactic
properties.  
For example, galaxy mass measurements, based on
either the virial theorem or Jeans equation modeling, depend linearly 
on the galactic radius \citep[e.g.,][]{vandermarel94a,wolf10a}.
While estimating structural properties should be straightforward,
practical considerations such as a limited field size or background
estimation can make this a complicated problem.

We have entered a new regime in estimating structural parameters with
the discovery of Ultra-Faint Dwarf (UFD) galaxies around the Milky Way
and M31 \citep{willman05b,willman05a,zucker06a,zucker06b,
  belokurov06a,belokurov06b,belokurov10a, martin09a}.  These objects
are found via statistical over-densities of resolved stars in the
Sloan Digital Sky Survey (SDSS) and other large imaging surveys with the
faintest ones having total luminosities of merely $M_V \sim
-1.5$, or $300$\,L$_{\sun}$ \citep{martin08a,geha09a,belokurov10a}.  The exact
nature of an individual UFD is often debated (is it a star cluster or dark
matter-dominated galaxy? is it in dynamical equilibrium?) 
and structural parameters are an important tool in this
assessment.  However, at the detection limits of the SDSS, their low
total luminosity translates into a few tens to hundreds of individual
stars observed as UFD members.  In this regime, a few stars can
significantly alter the derived size or total luminosity of a given object.

Structural parameters are determined throughout the literature by
measuring the surface brightness of a galaxy as a function of radius,
and then fitting an analytic function to this profile
\citep[e.g.,][]{caon93a}.  The surface brightness profile is usually
well determined for galaxies based on an integrated light profile, or
in the case of nearby luminous ($M_V < -8$) Local Group galaxies, by
summing many thousands of stars in concentric annuli \citep{irwin95a}.
In applying these techniques to the UFD galaxies, shot-noise due to
the low number of available stars can produce unreliable results.  As
an alternative to surface brightness fitting, \citet{martin08a} determined
structural parameters for the Milky Way UFD galaxies using a Maximum
Likelihood (ML) method, building on previous work by \citet{kleyna98a}
and \citet{westfall06a}.  In this method, a fixed analytic profile
is assumed and its free parameters (typically galaxy center,
half-light radius, ellipticity, position angle and the background
level) are fit simultaneously using all the available stars
\citep[see also ][]{sand09a,munoz10a}.  This avoids the need to bin or 
smooth data and should provide more reliable estimates for systems with low
number of stars.

There are regimes, however, where even the maximum likelihood method
will produce unreliable structural parameters.  
In this paper, we explore the
transition region between the extreme case where only
a few member stars of a resolved galaxy are identified 
and the well measured
galaxies as described above.  A quantitative understanding of this
regime will be of increasing importance since large surveys, such as
SkyMapper,
Pan-STARRS, DES, and LSST, are predicted to discover many tens to
hundreds more low luminosity systems \citep{tollerud08a}.

Here we perform a representative set of simulations to explore the
accuracy of the ML method for structural parameter
estimation in cases when the number of stars detected in a galaxy is a
few thousand or less.  
In \S\,\ref{ssec:sims} and \S\,\ref{ssec:ml} we describe the grid of
simulated resolved galaxies, and the ML method used to determine structural
parameters in these simulations.  
We then consider parameter
optimization for the particular cases of the SDSS and LSST
surveys(\S\,\ref{sec:surveys}), as well as targeted observations with
the Canada-France-Hawaii-Telescope (CFHT, \S\,\ref{sec:targeted}).
In \S\,\ref{sec:results} we explore
the accuracy of the recovered structural parameters split into the
separate effects of density contrast, the total number of stars, and
the effects of a limited field-of-view (FOV).  
Finally, in \S\,\ref{sec:summary} 
we summarize our results to
distill them into a set of generalized guidelines for future observations.

\section{Methods}

How well can structural parameters be measured in the regime where
a galaxy is detected as only a low number of resolved stars?  To answer
this question, we generate a series of simulated UFD galaxies and
assess how different observational variables impact our ability to
determine their structural parameters.  
We generate a series of
simulated resolved galaxies, varying the total number of stars, the
relative density of background stars and the physical size of the
object relative to the FOV.  We then determine the
structural parameters of these systems using a maximum
likelihood method and compare our recovered values back to the true inputs.

\begin{figure}
\plotone{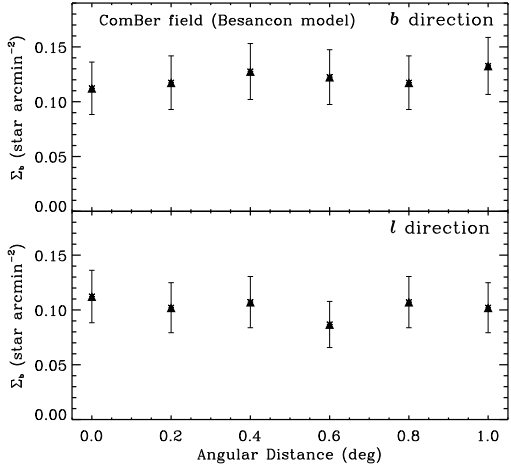}
  \caption{Variation of the stellar background density as a function of both Galactic
latitude and longitude at the position of the ComBer satellite over a $1^{\circ}$ region.
These data were simulated using the Besan\c{c}on Milky Way model. Variations in the
background counts are consistent with Poisson noise (error bars).
}
  \label{fig:besancon}
\end{figure}

\subsection{Simulations}
\label{ssec:sims}

The stellar density profiles of dwarf galaxies, including the UFDs,
are typically modeled using analytic density laws such as
exponential, Plummer \citep{plummer11} or King \citep{king62a}
profiles.  We construct our model galaxies using a Plummer profile,
but note that this choice does not affect our conclusions provided the 
same density law is used to generate both the simulations and recover 
the structural parameters. The projected Plummer density profile is:

\begin{equation}
\label{eq:plummer}
  \Sigma_{\rm dwarf}(x,y) = \Sigma_{0} \left[1 + \left(\frac{(x-x_{0})^{2}+(y-y_{0})^{2}}{R_{1/2}^{2}}\right) \right]^{-2} \,\,,
\end{equation}

\noindent where $R_{1/2}$ is the half-light radius, the galaxy center is at 
($x_{0}$, $y_{0}$) and $\Sigma_{0}$ is the central density.
As we are interested in exploring the effects of a limited
FOV on the recovered structural parameters, we 
measure $R_{1/2}$ in units of FOV, $L$.

Any stellar survey will also include some contaminating sources,
either foreground Milky Way stars or unresolved background galaxies.
We place our satellites over a 
homogeneous background density $\Sigma_{b}$ with Poisson variations.
Throughout this paper, we assume $\Sigma_{b}=1$ over the relevant area,
and report the central density $\Sigma_{0}$ in units of the background 
density. We have assessed the validity of our assumption of a homogeneous
background using both the Besan\c{c}on model at the Galactic position of
Coma Berenices (ComBer) and Bo\"otes~II (Boo~II), and CFHT imaging of 
the same UFD galaxies. We find that
variations for both datasets, in bins of 12 arcmin across a one degree
patch in the simulations and the one degree images, are on
the order of 10\% and are consistent with Poisson noise 
(see Figure~\ref{fig:besancon}). We conclude
that background variations over an area of one degree, relevant for
Milky Way UFDs, are not significant and our choice of a homogeneous
background, as a first approximation, is justified.

Given the Plummer profile above and assuming circular symmetry, a
simulated dwarf galaxy is fully described by three free parameters:
$R_{1/2}$, $\Sigma_0$ and $N_T$.  The final parameter $N_T$ is the
total number of objects, including UFD stars and background detections.  
In Table~\ref{tab:surveys}, we list approximate values for a few 
representative dwarf galaxies as observed by different instrument/survey 
combinations.

\begin{deluxetable}{lcccccc}
\tablecaption{
Approximate value for our different parameters for three representative dwarf galaxies 
    assuming SDSS and LSST-like surveys as well as targeted observations with MegaCam on the CFHT.
   We also list the assumed $r$-band depths and FOVs.} 

\tablehead{
\colhead{Survey/Telescope} & \colhead{$r$-band depth} & \colhead{$L$}  & \colhead{$R_{1/2}$} & \colhead{$N_{\rm dwarf}$} & \colhead{$N_{T}$\tablenotemark{a}} & \colhead{$\Sigma_{0}$}
}
\startdata
    \hline
    \multicolumn{5}{l}{ComBer ($M_V = -3.8$, $R_{1/2}=6.0\arcmin$, d = 44\,kpc)\tablenotemark{b,c}:} \\
    SDSS\tablenotemark{c}            & 22.5  &  $2^{\circ}$ &  0.05     &  99  & 3310  & $\sim5$    \\
    LSST$_{\rm single-exp}$  & 24    &  $1^{\circ}$ & 0.1      &  560  & 1305  & $\sim25$   \\
    LSST$_{\rm co-added}$    & 27    &  $1^{\circ}$ & 0.1      &  3420 & 8605  & $\sim22$ \\
    CFHT                     & 24.5  &  $2^{\circ}$ & 0.05     &  750  & 3740  & $\sim35$    \\
    CFHT                     & 24.5  &  $1^{\circ}$ & 0.1      &  750  & 1500  & $\sim35$    \\
    \hline
    \multicolumn{5}{l}{Segue~I ($M_V = -1.5$, $R_{1/2}=4.4\arcmin$, d = 23\,kpc)\tablenotemark{c}:}\\
    SDSS\tablenotemark{c}            & 22.5  &  $2^{\circ}$ & 0.037    &   65   & 4340  & $\sim3$      \\
    LSST$_{\rm single-exp}$  & 24    &  $1^{\circ}$ & 0.075    &   200  &  945  & $\sim16$     \\
    LSST$_{\rm co-added}$    & 27    &  $1^{\circ}$ & 0.075    &  1060  & 3610  & $\sim25$ \\
    CFHT                     & 24.5  &  $2^{\circ}$ & 0.037    &   260  & 3245  & $\sim21$     \\
    CFHT                     & 24.5  &  $1^{\circ}$ & 0.075    &   260  & 1005  & $\sim21$     \\
    \hline
    \multicolumn{5}{l}{Boo~II ($M_V = -2.7$, $R_{1/2}=4.2\arcmin$, d = 42\,kpc)\tablenotemark{c}:}\\
    SDSS\tablenotemark{c}            & 22.5  &  $2^{\circ}$ & 0.035    &   37   & 3940  & $\sim3$    \\
    LSST$_{\rm single-exp}$  & 24    &  $1^{\circ}$ & 0.07    &   210  &  960  & $\sim20$     \\
    LSST$_{\rm co-added}$    & 27    &  $1^{\circ}$ & 0.07    &  1280  & 6020  & $\sim17$ \\
    CFHT                     & 24.5  &  $2^{\circ}$ & 0.035    &   290  & 3275  & $\sim25$     \\
    CFHT                     & 24.5  &  $1^{\circ}$ & 0.07    &   290  & 1035  & $\sim25$     \\
    \hline
\enddata
  \tablenotetext{a}{The number of background counts has been derived using background density 
  values taken from \citet{martin08a} for 
  SDSS and from \citet{munoz10a} for the CFHT/MegaCam survey. 
  We note that in general, the latter values are lower than those from SDSS, despite the deeper 
  photometry. 
  For the case of an LSST-like survey we used 
  CMD-filtered counts from the Hubble Ultra Deep Field
  and assumed that $90\%$ of background galaxies can be removed.}
  \tablenotetext{b}{From \citet{munoz10a}.}
  \tablenotetext{c}{From \citet{martin08a}.}
  \label{tab:surveys}
\end{deluxetable}

\begin{figure}
\begin{center}
  \includegraphics[width=0.42\textwidth]{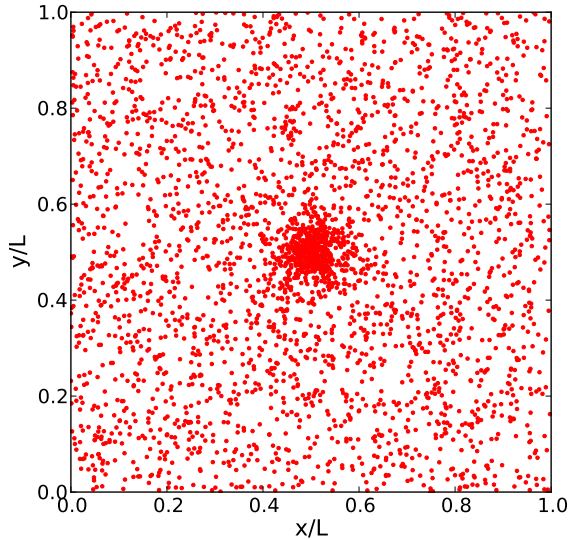}
\end{center}
  \caption{Simulated star map of an ultra-faint galaxy with
    parameters: $R_{1/2}=0.07$, $\Sigma_{0}=25$ and $N_{T}\sim1000$, 
    similar to a Boo~II-like satellite observed with the CFHT/MegaCam
    combination down to a magnitude limit of $r=24.5$.
    }
  \label{fig:boo2-cfht}
\end{figure}

\begin{figure}
\begin{center}
  \includegraphics[width=0.42\textwidth]{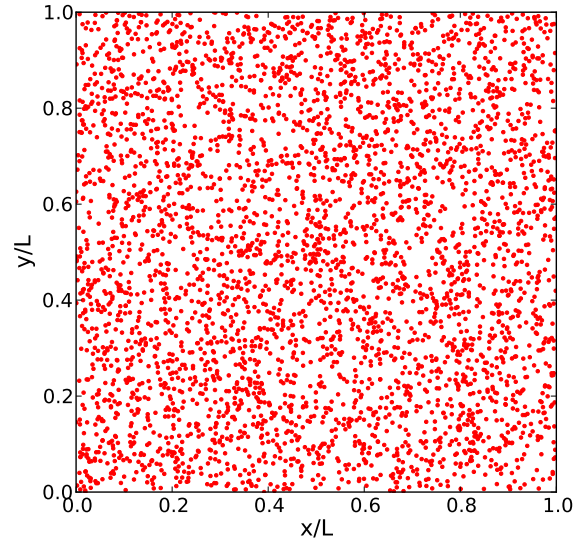}
\end{center}
  \caption{Same as Figure \ref{fig:boo2-cfht}, but for a Boo~II-like
    satellite as observed in the SDSS. This realization has
    parameters: $R_{1/2}=0.035$, $\Sigma_{0}=3$ and $N=4000$.  While
    the galaxy is not obvious by eye, it is detected as a statistical
    overdensity of star.}
  \label{fig:boo2-sdss}
\end{figure}

In Figures~\ref{fig:boo2-cfht} and \ref{fig:boo2-sdss} we show a dwarf galaxy 
simulated under different observing conditions.  Figure~\ref{fig:boo2-cfht} 
shows one realization of a satellite resembling the UFD galaxy Bo\"{o}tes~II 
(Boo~II, $M_V= -2.7$, $d=42$\,kpc) as it would be observed by the CFHT/MegaCam combination down to 
a magnitude limit of $r=24.5$.  The galaxy is clearly seen above the background 
level as a centrally concentrated group of resolved objects.  In contrast,
Figure~\ref{fig:boo2-sdss} shows the same galaxy as observed by the
SDSS.  Due to the much shallower data, the galaxy is difficult to
discern above the background counts.  Indeed, the UFD galaxies were
only recently discovered because the stellar density contrast between
the background and member stars is low-- in a typical image a UFD
cannot be ``seen" without additional filtering of the background.

We run a grid of simulations to cover most of the range of UFD observations presented 
in Table~\ref{tab:surveys}.  We vary $R_{1/2}$ between 0.025 and 0.5, $\Sigma_{0}$
from 3 to 100 and the total number of stars $N_{T}$ from 300 to 5000. 
For 
simplicity, the dwarf galaxies are placed at the center of the simulated field.  

\subsection{The Maximum Likelihood Method}
\label{ssec:ml}

We estimate the structural parameters of our simulated galaxies using a maximum 
likelihood (ML) method. We assume that the stars are distributed according to a 
density profile,

\begin{equation}
\Sigma (x, y, {\mathbf p}) = \Sigma_{\rm dwarf}(x, y, {\mathbf p}) + \Sigma_b
\end{equation}

\noindent where $\Sigma_{\rm dwarf}$ is the Plummer profile of equation 
\ref{eq:plummer} and $\Sigma_{b}$ is an arbitrary background density we 
choose to normalize to one.  The density depends both on the position in the 
field $(x,y)$ as well as the structural parameters ${\mathbf p} = (R_{1/2},
\Sigma_{0}, x_{0}, y_{0})$. Note that, in addition to the half light radius 
and central surface density, we also fit for the central position of the galaxy 
$(x_0, y_0)$. We continue to assume circular symmetry for simplicity.

The probability ${\cal L}_i$ of observing star $i$ is then given by

\begin{equation}
{\cal L}_i = \frac{ \Sigma(x_i, y_i, {\mathbf p})}{\int_{\Omega}\Sigma(x,y,{\mathbf p}) dx\,dy}
\end{equation}

where $\int_{\Omega}$ is the integral over the observed field and normalizes the 
probability distribution. The probability of observing the full data set given a 
particular choice of parameters ${\mathbf p}$ is therefore

\begin{equation}
{\cal L} = \prod_{i} {\cal L}_{i} \,.
\end{equation}
The maximum likelihood estimate of the parameters is defined to be the vector 
${\mathbf p}$ that maximizes the probability of observing the data.
The initial values of the measured parameters were chosen arbitrarily
(not at the true value); changing the starting position made no
difference to the final results.

Comparing this to previous work \citep{martin08a,sand09a,munoz10a}, we note that 
the probability is invariant under multiplicative scalings, explicitly justifying
our fixing $\Sigma_{\rm b}=1$. An alternative approach is to explicitly fit for 
the background density, in which case the normalization integral is then the 
total number of stars $N_{\rm T}$. This is the approach taken by 
\citet{martin08a}, although they make a further simplifying assumption that the
FOV is large enough that integrating the dwarf profile over
the field is equivalent to integrating it to infinity.
We do not make this assumption here, since we are also interested in
cases where the FOV is comparable to the size of the dwarf.

For each set of free parameters ($R_{1/2}$, $\Sigma_0$ and $N_T$) we
run 1000 realizations and generate a distribution for each of the
recovered inputs.  To assess the quality of the recovered parameters,
we report the width of these distributions.  We define the quantity
$\sigma_{\bf p}$ as the half width of the central 68\% of the
distribution.  While the distributions are not necessarily Gaussian,
this quantity is equivalent to the one-sigma width for a Gaussian
distribution. In addition, it is worth mentioning that in all cases,
no appreciable bias in the recovered parameters is observed, that is,
the observed distribution of measured values is roughly centered 
on (within at most 10\% of) the true input number.
 In the analysis presented below, we focus primarily on
the behavior of the half-light radius and therefore report only
$\sigma_{R_{1/2}}$ values.

\section{Studying UFDs in Wide-Field Imaging Surveys}
\label{sec:surveys}

In studying UFD galaxies, two different scenarios are relevant:
observations made using surveys, and targeted observations with a specific 
telescope/instrument combination.  These two cases present different observational
constraints and we examine each separately.

All of the UFDs have so far been discovered in the SDSS, but this
sample is far from complete due to the survey magnitude limits and sky
coverage.  Upcoming surveys, such as SkyMapper, Pan-STARRS, DES and
LSST, are expected to find many tens to hundreds more satellites
\citep{tollerud08a}.  
For any wide-field imaging survey two variables pertinent to
estimating structural parameters are fixed: the number of observed
stars belonging to the dwarf galaxy $N_{\rm dwarf}$, and the density
contrast. Both depend on the quality and depth of the photometry.  For
a given satellite's angular size and distance, the central density
depends directly on the number of member stars detected by the survey,
and is set primarily by the depth of the photometry, and to
a lesser extent, by the quality of the data.  
The background density increases with the depth of the survey but better 
quality data allows for more efficient removal of background unresolved 
galaxies.  The only free choice an observer faces is the area around the 
satellite used to carry out the structural parameter determination.  This 
choice will impact both the relative size of the satellite with respect to 
the FOV and the total number of stars $N_{T}$.
Na\"{i}vely, one expects the 
area around the satellite to have a mild effect, if any, on the accuracy of 
the derived structure, but in the new regime of ultra-faint satellites it
is an essential observational consideration.

We focus below on two surveys, SDSS and LSST, and examine how
selecting different area sizes affects the accuracy of the measured
half-light radii.

\begin{figure}
\plotone{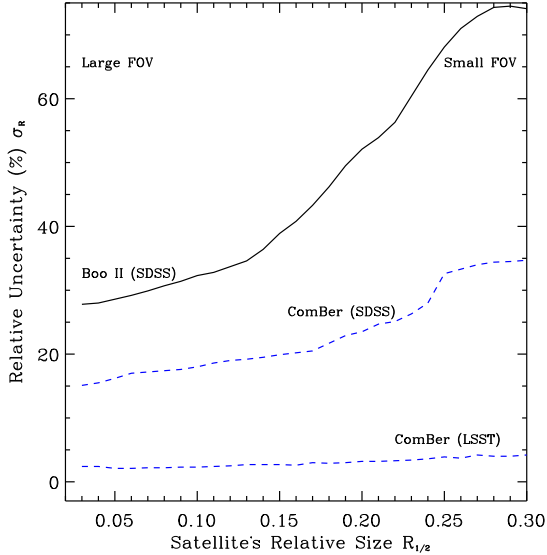}
  \caption{Fractional uncertainty $\sigma_{R_{1/2}}$ as a function of the satellite's 
  relative size, $R_{1/2}$ for two UFDs, Boo~II (solid lines) and ComBer (dotted lines), 
  ``observed" with two photometric surveys, SDSS (upper lines) and LSST (lower line).  
  For each line in this figure, the density contrast has been kept constant and therefore 
  the total number of stars varies from high numbers on the left of the lines to low 
  numbers on the right.  This figure shows that when the density contrast is fixed, 
  higher values of $R_{1/2}$ will not necessarily result in better uncertainties because 
  the number of stars decreases rapidly.
  }
  \label{fig:sdss_lsst}
\end{figure}

\begin{figure}
\plotone{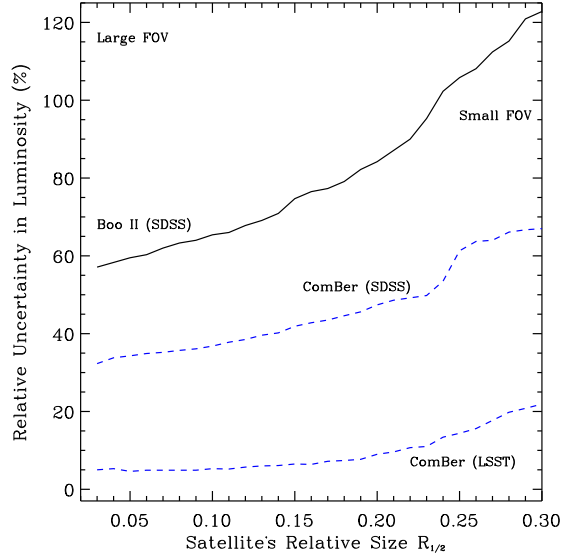}
  \caption{Same as Figure \ref{fig:sdss_lsst}, but for the fractional uncertainty 
  in luminosity $\sigma_{L}$}
  \label{fig:sdss_lsst_lum}
\end{figure}

\subsection{SDSS}
\label{subsec:sdss}

To illustrate how our simulations can guide the data analysis for
study of UFDs, we carry out mock observations for two example dwarf
galaxies: Boo~II \citep[$M_{V}=-2.7\pm0.9$, $l=353^{\circ}.7$, $b=68^{\circ}.8$;][]{martin08a} and Coma Berenices
\citep[ComBer, $M_{V}=-3.8\pm0.6$ $l=241^{\circ}.9$, $b=83^{\circ}.6$;][]{munoz10a}.  The former is one of the smallest
and faintest of the UFDs, while the latter is rather ``typical''.  We
use for reference the structural parameters of \citet{martin08a} who
used SDSS data to carry out a ML study of all UFDs discovered before 2008.  
For their structural analysis, these authors selected stars within $1\deg$ 
radius of each galaxy down to the SDSS magnitude limit of $r\sim22.5$.

From this catalog, Boo~II has a measured half-light radius of
$4.2\arcmin^{+1.1}_{-1.4}$ and is located at a distance of $42$\,kpc. 
In terms of the variables studied by our simulations, this results
in $\Sigma_{0}\approx3$, $R_{1/2}=0.035$ and $N_{T}=3940$.  
The simulations show that this combination of parameters for Boo~II yields
a $\sigma_{R_{1/2}}$ of roughly $30\%$ (similar to the uncertainty
reported by \citealt{martin08a}) and a 
$\sigma_{L}\sim90$, 
indicating that the measurement based on SDSS data
is highly inaccurate.  It is worth pointing out that \citet{walsh08a}, using 
deeper photometric observations of Boo~II  by roughly one magnitude, 
reported a $R_{1/2}=2.5\arcmin\pm0.8$, 
nearly $40\%$ smaller than the $4.2\arcmin$ found by \citet{martin08a}, 
consistent with the uncertainty obtained from our simulations.  
For the case of ComBer, the results of \citet{martin08a} give a half-light 
radius of $6.0\arcmin\pm0.6$ and a distance of $44$\,kpc, which results in
$\Sigma_{0}\approx5$, $R_{1/2}=0.05$ and $N_{T}=3310$.  This combination of parameters 
gives $\sigma_{R_{1/2}}\sim20\%$ and ,
$\sigma_{L}\sim50$, 
an improved value as compared to Boo~II.
In the case of ComBer, the uncertainties reported by 
\citet{martin08a} in their measured half-light radii are somewhat lower that those from 
our simulations. To estimate their errors, these authors assumed that the
likelihood function is well-described by a Gaussian distribution 
near its maximum, an assumption that probably breaks for the fainter objects.

As mentioned above, potential improvements in the accuracy of the measured 
sizes can only be achieved by changing the area around the satellite used
to select UFD member candidates.
In particular, we explore the effect of choosing a smaller region.
Given that the density contrast will remain constant, using a smaller FOV
will bring down the total number of stars.
To illustrate the trade-off between increasing $R_{1/2}$ while simultaneously
decreasing $N_{T}$, in Figure \ref{fig:sdss_lsst} we show the fractional 
uncertainty as a function of relative size for both Boo~II and ComBer 
when the background density and $N_{\rm dwarf}$ are derived from SDSS (upper lines). 
In addition, in Figure \ref{fig:sdss_lsst_lum} we show the fractional uncertainty in 
luminosity, which roughly correlates with $3\times \sigma_{R_{1/2}}$
We stress that while the measured uncertainties are shown as function of the 
satellite's relative size, the total number of stars is also changing. 
In particular, $N_{T}$ varies from higher values on the left to lower values
on the right.
This figure shows that better results are obtained for larger FOV,
with the decrease in uncertainty slowly plateauing as the FOV increases.  
This improvement is
a consequence of the increase in the total number of stars (see also \S5). 
In practice however, for regions much larger than those used by \citet{martin08a}
the improvements can be further limited, due to potential background fluctuations 
and gradients across the field.
We therefore conclude that it is not possible to improve substantially
the results of \citet{martin08a} by choosing a smaller FOV.
The ultimate limiting factor for a SDSS-like survey 
is the very low density contrast of the satellites 
with respect to the background, a reflection of both their low surface brightness and 
the relatively shallow limit of SDSS, a limitation that can be applied also
to SkyMapper, expected to reach a similar depth than SDSS.

In summary, for SDSS-like surveys, our simulations suggest 
that large FOVs will yield more accurate results, as long as the
background density remains homogeneous across the field.

\subsection{LSST}
\label{subsec:lsst}

The Large Synoptic Survey Telescope (LSST) will image the full sky
every two nights to a single exposure depth of $r=24$ over a period of
ten years, with a final co-added magnitude limit of $r\sim27$ \citep{ivezic08a}.
Because LSST is expected to find UFD satellites out to much larger
distances as compared to SDSS, the same issues will apply in regard to
measuring structural parameters in the limit of low numbers of member
stars.  An added challenge at LSST magnitude limits will be unresolved
background galaxies which will increasingly contribute to the background 
source counts fainter than $r>22$ (e.g., see Fig. 1 of \citealt{reid93a}).  
To assess the expected performance of 
LSST we carry out mock observations of the UFD galaxy ComBer, placed at 
two different distances: its true distance of 44\,kpc and farther away at 400\,kpc.

\begin{figure}
\plotone{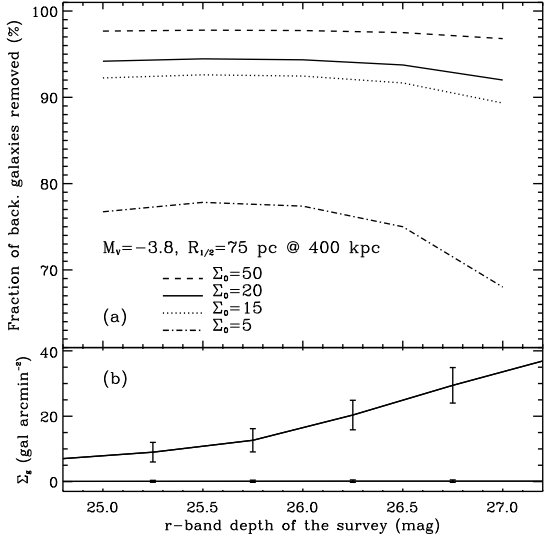}
  \caption{{\it Upper panel:} required fraction of background galaxies that must be removed as a 
  function of $r$-band depth to achieve three different density contrast values: 
  5, 15, 20 and 50.  
  This particular set of curves represents simulated data for a co-added LSST-like 
  survey of a satellite with absolute magnitude $M_{V}=-3.8$, half-light radius of 
  $R_{1/2}=75$\,pc located at $400$\,kpc.
  {\it Lower panel:} The upper line shows the number density of filtered background 
  galaxies used in panel (a) as a function
  of $r$ magnitude. Galaxy counts were obtained from the HUDF. As a comparison, the lower
  line shows count densities for Milky Way stars calculated using the TRILEGAL tool.
   }
  \label{fig:gal_star}
\end{figure}

We estimate the number of UFD stars, $N_{\rm dwarf}$, detectable down to
the full co-added LSST survey magnitude limit using a theoretical luminosity
function typical of the UFDs so far discovered: a 13\,Gyr old
population with overall metallicity of [Fe/H]$=-2.27$
\citep{girardi04a}, assuming a \citet{chabrier01a} initial mass
function.  
For ComBer at 44\,kpc, we obtain $N_{\rm dwarf} = 3420$. We combine this
information with the known angular size of this system to determine
its central density.  To derive a density contrast we then need to estimate
the background density.  For a survey like LSST, this variable is
ultimately tied to our ability to characterize the 
contamination, due either to foreground Milky Way stars or unresolved
background galaxies.  We estimate the expected contamination from
Milky Way stars at these magnitudes by generating fake photometric
data for our Galaxy using the TRILEGAL tool \citep{girardi05a}. For
the relevant magnitude range ($22<r<27$) this yields foreground star
densities that are negligible compared to those of background galaxies
within our Color-Magnitude-Diagram (CMD) 
window\footnote{We define a region 
around the main sequence of the isochrone described above, located at 44\,kpc.}
(as we illustrate in Figure~\ref{fig:gal_star}(b)).
Our primary contamination will then come from
unresolved background galaxies.  To estimate their contribution to
$\Sigma_{b}$, we use the Hubble-Ultra-Deep-Field (HUDF) catalog
under the assumption that the distribution of galaxies is isotropic.
Using a region of the CMD similar to the one used to select UFD stars, 
we estimate a background density of $\Sigma_{\rm gal}=14.4$\,gal\,arcmin$^{-2}$ 
down to the limit of $r=27$.  This yields $\Sigma_{0}=2.2/(1-\mu)$, where
$\mu$ is the fraction of background galaxies that can be potentially
removed photometrically.  If we assume, for instance, that we can
remove 80\% of this contamination\footnote{This assumption is based on
  the results of \citet{munoz10a} where we found that using the
  DAOPHOT morphological parameters $\chi$ and sharp, we can remove up
  to 80\% of background galaxies down to a magnitude limit of
  $r\sim24.5$. In practice, this fraction is likely to be lower for
  $r=27$.}  
we obtain
$\Sigma_{0}\sim11$.
This number may seem surprisingly low compared to the density contrast
achieved by \citet{munoz10a} using shallower data.
However, this is only a reflection of the fact that unresolved galaxy counts 
grow significantly faster at faint magnitudes. 
This is true even if we allow the density of background galaxies to vary by 
a factor a few with respect to the HUDF. 

A benefit of deeper photometry is the increase in 
the total number of detections $N_{T}$. 
For ComBer at 44\,kpc, and assuming we can remove 80\% of the galaxy counts, 
we get $N_{T}=5050$ for a half-light radius to FOV ratio of $0.25$, yielding $\sigma_{R_{1/2}}\sim4\%$.
The lower line in Figure~\ref{fig:sdss_lsst} shows the dependence of the relative 
uncertainty with the satellite's relative size for this particular choice
of density contrast value.
Both the density contrast and the total number
of stars are large enough that the measured uncertainty is rather
insensitive to the choice of FOV.

One of the unique aspects of LSST related to UFDs is the expectation of finding 
galaxies as faint as $M_{V}=-1$ out to the virial radius of our Galaxy, located
roughly at 400\,kpc \citep{tollerud08a}.
In Figure~\ref{fig:gal_star}(a)
we show the 
required fraction of background galaxies that must be removed as a function of 
$r$-band depth to achieve a given density contrast in the context of observing 
a ComBer-like satellite ($M_{V}=-3.8$, $R_{1/2}=75$\,pc)
located at 400\,kpc. 
The figure shows curves for three different values of $\Sigma_{0}$, 15, 20 and 50.
To achieve a density contrast of at least 20 for magnitudes 
fainter than $r=25$, more than 90\% of the unresolved galaxies must be subtracted. 
We note that while the fraction of galaxies that need to be removed does not
increase at deeper magnitudes, the actual background density is a strong function
of depth 
(Figure~\ref{fig:gal_star}(b)).

Alternatively, one can ask what is the 
limiting distance at which we can characterize the half-light 
radii of a given UFD with an intrinsic uncertainty of 10\% or better (or
equivalently, reach a density contrast of 20 or higher).
The answer to this question is ultimately a function of our ability
to separate efficiently stellar from non-stellar detections.
For illustration purposes, we will assume an upper limit of 90\% of
galaxy removal\footnote{This is in essence an arbitrary number at this time, since 
  we do not know with certainty what will be possible to achieve with LSST, but we deem 
  it a sufficiently challenging upper limit for the purpose of setting a limiting distance.}. 
With this number we find that for a ComBer-like satellite, we obtain a density
contrast of 20 or more out to about 400\,kpc, but not farther.
In contrast, for fainter UFDs,
i.e. with $M_{V}\sim-2$,
comparable to systems like Segue~I, Boo~II or Willman~I, the limit distance is
$200-250$\,kpc. The same is true for brighter ($M_{V}\sim-5$) but larger 
UFDs (i.e. systems with half-light radii larger than 100\,pc).
Achieving density contrasts that allow proper determination of structural parameters 
for distant UFDs will thus present a major observational challenge.

\section{Targeted Observations of UFDs}
\label{sec:targeted}

Follow-up targeted imaging of UFDs can provide more accurate structural
parameters than the discovery survey data
\citep{sand09a,munoz10a}.  We next explore how to optimize targeted
observations to provide the most accurate structural parameter
estimates.  In contrast to survey data, the practical observing
constraint is the total exposure time which translates into
a trade-off between photometric depth versus area covered.  
We choose to analyze a ComBer-like satellite, although the conclusions 
are applicable to the majority of the UFDs,
for two scenarios 
requiring the same exposure time: a single $1\times1\deg^{2}$ 
pointing down to $r=24.5$ and four pointings down to 
$r=23.0$. This choice of area was motivated by \citet{munoz10a} 
who used the telescope/instrument combination of CFHT/MegaCam.

\begin{figure}
\epsscale{0.85}
\plotone{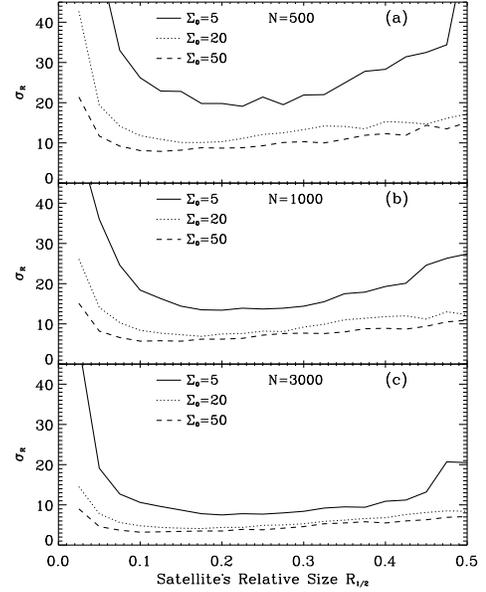}
  \caption{Fractional uncertainty $\sigma_{R_{1/2}}$ as a function of the
satellite's relative size, $R_{1/2}$. Different panels correspond to
different choices of the total number of stars $N_{T}$. In each panel,
three different density contrasts are shown.
This figure shows that (i) density contrasts higher than 20 do not
result in a significant improvement in the uncertainty of the
measured half-light radii and (ii) the optimal range FOV
size is such that $0.1<R_{1/2}<0.3$.
}
  \label{fig:twopanels}
\end{figure}

The FOV of the instrument is now fixed, and thus $R_{1/2}$
is determined by the size of the satellite.  For observations of ComBer
down to $r\sim24.5$, the extrapolated number of stars belonging to the
dwarf galaxy corresponds to $N_{\rm dwarf}\sim750$.  Assuming a
background density similar to that estimated by \citet{munoz10a}, we
obtain $N_{T}\sim1500$, $R_{1/2}=0.1$ and a density contrast of
$\Sigma_{0}\sim35$. Thus, for a single deep pointing covering $1\deg^{2}$ 
we obtain $\sigma_{R_{1/2}}\sim5\%$.
In comparison, observations down to $r=23$ covering twice the size of 
the FOV (four times the surveyed area),  result in 
$\Sigma_{0}\sim10$, $R_{1/2}=0.05$ and $\sigma_{R_{1/2}}\sim11\%$. 
This example shows that, in general, given the choice of depth versus
area, the former yields better uncertainties.
In light of this result, it is relevant to assess whether there is an
optimal depth for which uncertainties will be under 10\%.
Figure \ref{fig:twopanels} shows that, for a given total number of stars and
satellite's relative size, the overall uncertainties only marginally
improve for density contrasts beyond 20.  Thus, in planning targeted
observations of UFD galaxies, exposure times should be calculated so
that $\Sigma_{0}\sim20$.  Deeper photometry, presumably resulting in
greater density contrast, will not have an appreciable effect on the
overall quality of the measured parameters.  Of course, this does not
preclude the observer from seeking higher density contrasts for
different purposes.  
In addition, one should plan the area coverage so that,
ideally, $0.1<R_{1/2}<0.3$ and $N_{T}>1000$. Total number of stars as
low as $N_{T}=500$ will yield $\sigma_{R_{1/2}}\sim10-20\%$ within the
same range of satellite's relative size and density contrast value
(panel (a) of Figure \ref{fig:twopanels}).

\section{General Trends}
\label{sec:results}

The main goal of our simulations is to provide observers with a set of
tools to help select the appropriate set of parameters when designing
observations of UFDs.  
As shown in previous sections,
the distribution of recovered $R_{1/2}$ values will depend on all three
parameters simultaneously: total number of stars, satellite's relative size and
density contrast. However, it is possible to distill
these results into a set of ``rules of thumb" by isolating
the effects of each variable independently.

\begin{figure*}
\plotone{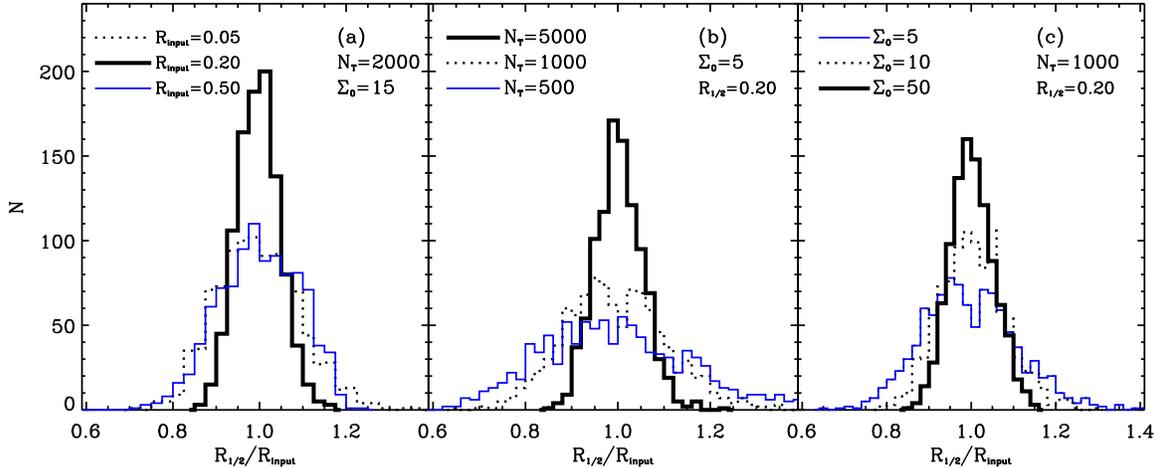}
  \caption{
Distribution of measured $R_{1/2}$ for simulations where we vary
one of the studied variables.
{\it Left panel:} 
we  vary relative sizes. 
The lowest value of $R_{1/2}=0.05$ corresponds
to a satellite occupying $10\%$ of the FOV, while the
highest value of 0.50 corresponds to a satellite covering the full FOV.
{\it Middle panel:} we vary the total number of stars. The density contrast 
and satellite's size are kept constant.
{\it Right panel:}
we vary the  density contrast. 
The total number of stars has been kept constant
at $N_{T}=1000$.
and $R_{1/2}=0.2$ 
   }
  \label{fig:effect_all}
\end{figure*}

In Figure 
\ref{fig:effect_all}(a)
we 
explore the effects of varying the satellite's relative size. 
In this figure, we show the distribution of $R_{1/2}$ for 
three cases where we have varied the size of the satellite.  For this particular 
example we fix $N_{T}=2000$ while simultaneously keeping the density contrast
fixed at $\Sigma_{0}=15$.  The general trend illustrated by panel (a)
is such that, for a constant number of stars, in the limit where the size of 
the satellite is comparable to the FOV ($R_{1/2}\sim0.5$) or when the field is too large 
($R_{1/2}<0.1$), our ability to recover the intrinsic value of the half-light 
radii decreases significantly.  
This is readily understood since for large satellite sizes, with respect
to the FOV, we are not
able to probe the full profile, while for small sizes, the number of UFD stars
decreases (for a fixed $N_{T}$) and the satellite's signal is diluted.
In practice however, as one uses a larger FOV, the number of stars
also increases, and therefore the measured uncertainty behaves
as shown in Figure \ref{fig:sdss_lsst}, that is, no improvement
or worsening in accuracy is achieved beyond $R_{1/2}\sim0.1$. 
Observing larger FOVs should not
be a problem, since one can always use only a subset of the data.

We next investigate the effect of varying the total number of stars,
$N_T$.  
Figure \ref{fig:effect_all}(b) shows the
distribution of recovered $R_{1/2}$ for three simulations with input
values $N_{T}=500$, 1000 and 5000, bracketing values obtained for
various current and future surveys (see Table \ref{tab:surveys}), and
constant $\Sigma_{0}=5$ and $R_{1/2}=0.2$.  The figure shows that when
all other variables are constant, increasing the total number of stars
improves the parameter estimation.  
For the entire grid of simulations we find that 
for a wide range of 
$\Sigma_{0}$ and $R_{1/2}$ values,
uncertainties better than 10\% require a minimum of $N_{T}=1000$.

Finally, we investigate the effects of varying the density contrast
$\Sigma_{0}$, on the recovered half-light radii. In
Figure \ref{fig:effect_all}(c), we show the distribution of measured
half-light radii for three different choices of central density (in
units of the background density), 5, 10, and 50 for satellites with
$R_{1/2}=0.2$ and $N_{T}=1000$.  
Figure~\ref{fig:effect_all} shows that, as
intuitively expected, as the stellar density contrast between the
satellite and the background increases, the distribution of measured
half-light radii becomes narrower and better defined (i.e., more
Gaussian) yielding a median value closer to the intrinsic value.  We
find that for a given combination of $N_{T}$ and $R_{1/2}$, 
the improvement in the measured uncertainties is most significant for
  $\Sigma_{0}<20$. For higher values of the density contrast,
$\sigma_{R_{1/2}}$ decreases only marginally.

\section{Summary and Conclusions}
\label{sec:summary}

With the discovery of
the UFDs 
we have entered
a new luminosity regime dominated by shot noise and where traditional
methods to estimate structural parameters no longer yield reliable 
results.
In this
article, we have carried out mock observations of a suite of simulated UFD
galaxies aimed at exploring 
how different observational choices impact our ability to extract
reliable structural information for these systems. In particular, we
have investigated the effects of the total number of stars in a given
survey, central-to-background density contrast and relative size of
the satellite with respect to the FOV.

Our mock observations allow us to propose a
set of simple ``rules of thumb" designed to help plan studies of UFDs
with existing and/or upcoming surveys as well as plan follow-up
observations:
(a) the ratio of the radius of the satellite relative to the FOV 
must be smaller than $0.3$,
(b) 
the exposure time should be calculated so that 
density contrast is $\sim20$. Higher numbers will result in marginal gains
in parameter uncertainties and 
(c)
for observations of UFDs with photometric surveys the density contrast
will be fixed by the survey parameters, and therefore the primary
observational choice is the area around the satellite. This should be
chosen so that, ideally, $N_{T}>1000$.

It is worth keeping in mind that these rules of thumb are derived for
the best case scenario, i.e., the choice of density profile faithfully
describes the real data, the satellite is perfectly circular, and
the background density is homogeneous throughout the FOV.
In practice, all three assumptions are likely to be broken in varying
degrees, which will lead to further uncertainties not quantified by 
this study. This leaves room for many possible extensions to the present 
study. Future work should include, for instance, an exploration of the 
effects of recovering structural parameters using a density law different 
from that used to construct the simulations.  Including intrinsic elongation 
for the simulated satellites is also warranted given the high ellipticity 
found for most UFDs, as it is extending the analysis to satellites that 
present small tidal disturbances.

We have shown that the new regime of extreme luminosities 
probed by the UFD population often results in unreliable structural
parameters, especially when relatively shallow photometric surveys, such as SDSS, 
are used. In the particular case of the half-light radii, on which mass
estimates depend linearly, numbers based on SDSS catalogs can have intrinsic 
uncertainties as high as 100\% for the faintest and more diffuse UFDs.
These higher uncertainties will, for instance, 
propagate directly into the 
predicted $\gamma$-ray fluxes coming from UFDs due to dark
matter annihilation \citep[e.g.][]{strigari07b}. Assessing the reliability of the 
measured structural parameters must be regarded as critical in these studies.

We thank an anonymous referee for her/his useful comments.
We also thank Beth Willman and Josh Simon for useful discussion on this work.  R.R.M. and M.G.
gratefully acknowledge support by the National Science Foundation
under AST-0908752. 
M.G. also acknowledges support from the Alfred P.~Sloan Foundation.
R.R.M. acknowledges support from the GEMINI-CONICYT Fund, 
allocated to the project No. 32080010, and from CONICYT through projects 
FONDAP No. 15010003 and BASAL PFB-06.

\end{document}